# Enhancing tutoring systems by leveraging tailored promptings and domain knowledge with Large Language Models

Mohsen Balavar[1] and Wenli Yang[1\[0000-0002-4885-2531\]] and David Herbert[1\[0000-0003-1419-7580\]] and Soonja Yeom[1\[0000-0002-5843-101X\]]

[1]School of Information and Communication Technology, University of Tasmania .

**Abstract:** Recent advancements in artificial intelligence (AI) and machine learning have reignited interest in their impact on Computer-based Learning (CBL). AI-driven tools like ChatGPT and Intelligent Tutoring Systems (ITS) have enhanced learning ex-periences through personalisation and flexibility. ITSs can adapt to individual learning needs, and provide customised feedback based on a student's performance, cognitive state, and learning path. Despite these advances, challenges remain in accommodating diverse learning styles and delivering real-time, context-aware feedback. Our research aims to address these gaps by integrating skill-aligned feedback via Retrieval Aug-mented Generation (RAG) into prompt engineering for Large Language Models (LLMs) and developing an application to enhance learning through personalised tutor-ing in a computer science programming context. The pilot study evaluated a proposed system using three quantitative metrics: readability score, response time, and feedback depth, across three programming tasks of varying complexity. The system successfully sorted simulated students into three skill-level categories and provided context-aware feedback. This targeted approach demonstrated better effectiveness and adaptability compared to general methods.

## 1 Introduction

Computer-Based Learning (CBL) is a learning approach that has existed for decades [1]. Recently however, highly significant advances in AI through generative data and well-publicised and widely adopted approaches such as ChatGPT have shown large language models (LLM) can greatly enhance the learning experience [2], causing a re-newed interest in the potential impact of these technologies on CBL. One of the greatest advancements in recent years is the development of Intelligent Tutoring Systems (ITS) which can be highly adjustable and flexible to suit an individual's learning needs [3]. Existing ITS techniques utilise narrow AI [4] for individual learning support like iden-tifying learning gaps and generative AI to create personalised learning materials [3]. Narrow AI, also known as weak AI, is tailored for a limited or particular set of tasks. Unlike general AI, which aims to mimic human behaviour broadly, narrow AI is limited to its defined domain, and it lacks broader awareness. Despite these advancements, there are still gaps in customising ITS to suit various learning styles and provide real-time, context-aware feedback. A context aware ITS can personalise topics for individ-ual students by assessing their strengths and weaknesses, creating an accurate model of their cognitive state, learning pace, and requirements for an enhanced learning experience.

Creating an efficient learning portfolio is crucial for our system design, capturing a student's performance in the target domain and related areas in their learning journey. Our institution uses Intended Learning Outcomes (ILOs) to assess student progress in each subject, with every assessment item aligned to these outcomes for systematic evaluation. To address current shortcomings, our research focuses on building a responsive system that tailors feedback based on individual student portfolios and past performance. Using Retrieval Augmented Generation (RAG) [5], we retrieve relevant content from previous subjects to customise responses. We employ structured prompts to ensure precise, reliable, and thorough feedback, promoting skill enhancement in a structured manner. Our motivations drive us to systematically profile student skills, integrate skill-aligned feedback into Large Language Models (LLMs) for personalised tutoring to enhance learning with real-world scenarios and tasks. Here are our main contributions:

1. **Profiling Student Skills and Aligning with Learning Outcomes:** This involves continuously profiling student skills by evaluating their performance in prerequisite subjects and current progress in the target subject, ensuring alignment with predefined learning outcomes and facilitating targeted support.
2. **Integration of Skill-Aligned Feedback Prompts and Generated Knowledge into LLMs for Personalised Tutoring**: This integration facilitates the delivery of contextually relevant and finely tuned feedback to individual learners.
3. **Development and Evaluation of an Application for Enhanced Learning:** This entails the development by incorporating various real-world cases and tasks, as well as providing a comprehensive evaluation of its effectiveness.

## 2       Relevant Work

### 2.1     State-of-the-Art in Tutoring System Technologies

Intelligent Tutoring Systems (ITS) are sophisticated educational software solutions designed to deliver personalised instruction and feedback to learners. Leveraging the power of applied artificial intelligence (AI) techniques, these systems adapt to the unique needs and learning styles of individual students [3]. The primary goal of ITS is to enhance the learning experience by providing customised educational content, timely contextual feedback and guidance, ultimately leading to improved learning outcomes and student engagement. Table 1 summarises the current AI techniques utilised in intelligent tutoring system technologies.

Most current AI techniques used in ITS leverage narrow AI, which focuses on specific tasks such as predicting student learning outcomes, analysing student responses, and providing personalised feedback. However, they are often limited by the specificity and scope of their predefined tasks and rules. Generative AI, such as LLMs could be a transformative solution to these limitations by offering more versatile and dynamic capabilities. By incorporating generative AI, ITS can better simulate human-like interactions, understand a wider range of student inputs, and create more personalised and effective learning experiences.

**Table 1.** AI techniques commonly used in Intelligent Tutoring Systems (ITS).

| AI Technique | Function in ITS Systems | Limitations | Refs |
|---|---|---|---|
| Machine Learning/ Deep Learning | Extract features and make predictions from large datasets of student interactions and performance. | Requires large data for training, may struggle with interpretability, and can suffer from biases. | [6] |
| Natural Language Processing | Understand and interpret student responses to tailor system responses based on content and context. | Challenges in handling ambiguity and context-dependent meanings. | [7] |
| Knowledge Representation | Model domain knowledge and student misconceptions to generate personalised materials, explanations, and feedback. | May struggle with dynamically evolving content and context. | [8] [9] |
| Expert Systems | Facilitate domain expert human-like interaction between students and tutoring systems. | May lack the ability to handle uncertainties or novel situations. | [10] |
| Reinforcement Learning | Provide adaptive feedback and reward mechanisms to optimise strategies based on student engagement and progress. | May require significant exploration to find effective policies and can suffer from slow convergence and high variance. | [11] |

### 2.2 LLMs in Educational Technologies

LLMs have evolved significantly over the past decade, revolutionising the field of natural language processing (NLP). In education, LLMs have the potential to transform traditional teaching methods by providing personalised learning experiences, enhancing the accessibility of educational content, chatbots and virtual assistants, etc. The integration of LLMs into educational technologies is rooted in the ability of these models to understand and generate human-like text, which can be leveraged to support both teaching and learning processes. Table 2 summarises the current typical educational applications that utilise LLMs.

ITSs have evolved significantly due to the integration of LLMs. Traditionally, language models have been employed in procedural tutoring and problem-solving systems within ITS. This approach required meticulous engineering of prompts to ensure reliability and the reduction of LLM hallucinations. However, a more advanced method known as retrieval-augmented generation (RAG) [5] has emerged to enhance the capabilities of LLMs in ITS. RAG incorporates external text corpora into the LLM's output by retrieving relevant texts and making them available for the task at hand.

Table 2. Typical applications of Large Language Models (LLMs) in Education.

| AI Technique | Function in ITS Systems | References |
|---|---|---|
| Personalised Learning | GPT-3, GPT-4 | [12] [13] |
| Intelligent Tutoring Systems | GPT-3, GPT-4, BERT, RoBERTa | [14] |
| Automated Grading | BERT, GPT-3, GPT-4, | [15] [16] |
| Educational content creation | GPT-3, GPT-4, BERT, Gemini, T5 | [17] |
| Chatbots and Virtual Assistants | GPT-3, GPT-4, BlenderBot, Gemini | [18] |
| Formative Assessments | GPT-3, GPT-4, RoBERTa | [19] |

While LLMs with RAG offer significant potential in enhancing Intelligent Tutoring Systems, they also have some limitations.

*Reflection on Learning Outcomes:* RAG approaches may overlook learning outcomes, prioritising contextually appropriate responses over direct alignment with objectives, potentially limiting effectiveness.

*Consideration of Student Skill Levels:* RAG techniques may not account for individual skill levels, compromising personalised learning experiences by offering responses that aren't tailored to students' proficiency levels and needs.

## 3      Methodology

Our ITS system framework design consists of three main modules: (1) knowledge generation for student skills profiling, (2) developing tailored prompting mechanisms, and (3) integrating skill-aligned feedback prompts with generated knowledge into LLMs for personalised content generation. The combination of these modules involves leveraging the LLMs' capabilities to tailor educational materials to individual learners.

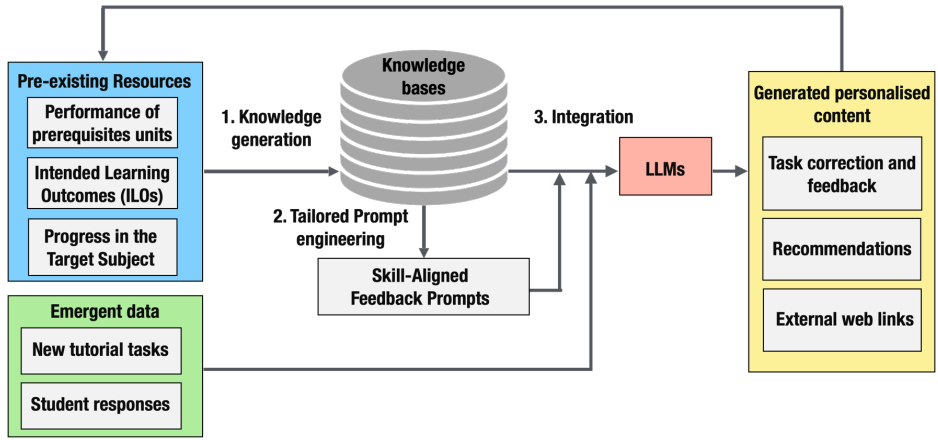

*Figure 1- The overall framework of the pilot ITS*

Enhancing tutoring systems by leveraging tailored promptings .. with Large Language Models

### 3.1 Knowledge Generation for Profiling Student Skills

To create a truly personalised experience, each student needs a knowledge base (called a *student's portfolio*) that can be used to tailor feedback with RAG ratings (Figure 1). This involves retrieving relevant information and structuring it for LLM prompt engineering, ensuring predictable LLM behaviour. The ITS will generate an individualised profile for each student in a target subject, tracking their performance in prerequisite subjects, individual learning objectives (ILOs), and progression in the target subject, including any assessments. These profiles also aid in planning future subjects and modules. Customised profiles enable the ITS to provide tailored support and guidance based on each student's unique needs and past performance.

Creating and maintaining students' profile-based learning progression aligns with prior course-level curriculum design and the definition of the target subject's ILOs. Each ILO is supported by assessment tasks with associated skills enhanced through specific tasks. By mapping these key skills and tasks to the relevant ILOs (via weekly tutorials and quizzes), we enhance the knowledge base's clarity, increasing the likelihood of retrieving relevant, contextual information. This approach ensures the LLM provides insightful, grounded feedback, reducing the chance of hallucinations.

In addition to previous subjects, we leverage completed assignments and projects in the target subject. Student performance is categorised into *below average*, *average*, or *above average* clusters, with feedback adjusted accordingly by the ITS. As the semester progresses, student portfolios become more comprehensive, allowing for increasingly tailored feedback. The quality of this knowledge base is crucial for precise, accurate, and relevant LLM feedback.

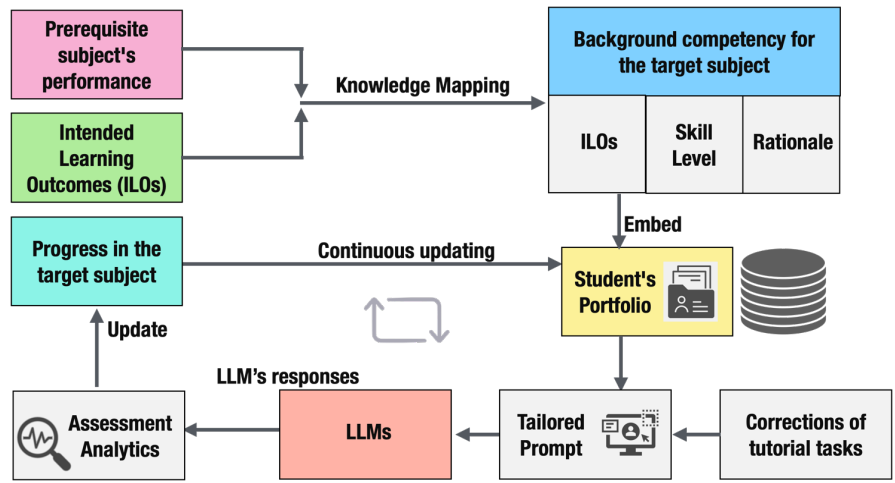

*Figure 2 - The process of creating a student's portfolio*

The pilot framework for creating a comprehensive knowledge base is illustrated in Figure 2. The knowledge base (*Student's Portfolio*) includes two main attributes: prior skills (performance in prerequisite subjects) and progress in the target subject throughout the semester. It also includes the ILO relationships between prerequisite and target

subjects – evaluating these relationships helps map presumed skills to current performance. This system ensures feedback aligns with students' categorised skill levels and competencies. Feedback can be analysed to track progression, providing continuous maintenance of the portfolio knowledge base.

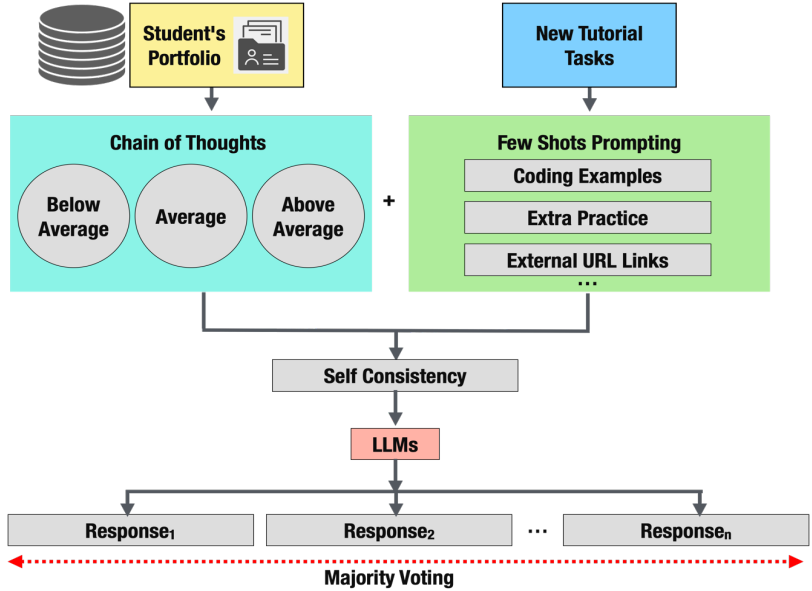

*Figure 3 - The prompting techniques in the pilot ITS*

### 3.2    Development of Tailored Prompting Mechanisms

Leveraging best practices in the field of prompt engineering allow a reduction in the bias of language models that might arise from their training data. This is especially important if the language models use reinforcement learning from human feedback (RLHF), which has a higher chance of introduced bias. When we systematically and logically instruct the language model through specialised prompt engineering, we can increase the probability of getting reproducibility and consistency in the model's output. By structuring the logical flow of questioning and answering, we can ensure that the model approaches a question in the same manner that a student would do. Understanding a process one step at a time and explaining the reasoning behind it can significantly improve the feedback provided. Three of the best practices that can be adopted in the ITS were tested and evaluated – they are *Chain of Thought*, *Few Shots Prompting* and *Self Consistency through majority voting* [20].

We provide multiple examples and reasoning steps of expected responses to the LLM to structure the output as requested in the ITS. By providing accurate feedback and examples, we aim to give the LLM a better understanding of the complexity of the task so that it can adapt better to any given scenario with a similar structure (Figure 3).

Enhancing tutoring systems by leveraging tailored promptings .. with Large Language Models

The *Self-Consistency* method is a technique used to obtain multiple answers for each task, and the system presents the final answer based on a majority vote over answers. This technique significantly enhances the performance of the ITS by utilising the most frequently occurring answer for the task at hand. It is crucial for maintaining consistency in providing feedback, especially for tasks where LLM might produce varying answers.

Table 3 shows our *Chain of Thoughts* prompt design for the students among skill ranking categories. Then multiple few shots prompting are embedded into our design. Table 4 shows a sample of *Few Shots* with the design of our proposed Chain of Thought. Finally, after multiple responses from the language model are generated for the same task, we use the *Majority Voting* technique to present the most relevant and consistent answer by utilising the *Self-Consistency* technique [21].

**Table 3.** Chain of Thought prompts design.

| **Below average** | **Average** | **Above average** |
|---|---|---|
| Identify and describe basic machine learning methods. | Distinguish between different algorithms. | Deeply understand the mathematical foundations behind each machine learning method. |
| Follow step-by-step instructions to use basic machine learning tools | Recommend appropriate libraries and techniques for specific tasks. | Encourage to integrate advanced techniques to implement. |
| Offer additional external links for more basic practices. | Provide additional external links on advanced machine learning topics. | Provide additional external links for continuous improvement in applying machine learning method. |

**Table 4.** Example of Few Shots with the design of our proposed Chain of Thought

**Task:** *Create a linear regression (LR) model using scikit-learn*.
**Student's Response:**
```
from sklearn.linear_model import LinearRegression
import pandas as pd
data = pd.read_csv('data.csv')
model = LinearRegression()
model.fit(X, y)
```

| | **Feedback** | |
|---|---|---|
| **Below average** | **Average** | **Above average** |
| *Basic description of LR* | *Difference between LR and classification* | *Mathematical formulae and description of LR* |
| 1. Import Libraries:<br>    *- Explanation*<br>    *- Coding examples*<br>2. Load Data:<br>    *- Explanation*<br>    *- Coding examples* | 1. Specific correction of missing splitting data<br>    *- Explanation*<br><br>2. Libraries and techniques recommendation: using techniques like `train_test_split` from *scikit-learn*.<br>    *- Coding examples* | 1. Better to consider data pre-processing.<br>    *- Explanation*<br>    *- Coding examples*<br>2. Better solution using k-fold cross-validation.<br>    *- Explanation*<br>    *- Coding examples* |
| 3. Split Data:<br>  *- Explanation*<br>  *- Coding examples*<br>4. Create and Train Model.<br>  *- Explanation*<br>  *- Coding examples* | - | 3. Integration using pipelines.<br>  *- Explanation*<br>  *- Coding examples*<br>- |
| Additional External Links for More Basic Practices:<br>  *- URL links* | Additional External Links for advanced ML topics:<br>  *- URL links* | Additional External Links for continuous improvement in applying machine learning method:<br>  *- URL links* |

## 4     Evaluation and Results

### 4.1    Data collection and preparation

The data used in this study is a synthesised dataset containing the marks of 30 students for two prerequisite subjects, *C108* and *C205*, and the marks for 2 weeks of tutorials and quizzes for the current subject, *C315*. The simulated data was generated to follow a normal distribution with a mean of 72 and a standard deviation of 8.

Assuming a pass mark of 50 for the prerequisite subjects, students were categorised as *below average* (marks between 50 and 65), *average* (marks between 65 and 80), and *above average* (marks above 80). Figure 4 shows a sample of the synthesised data used in this study.

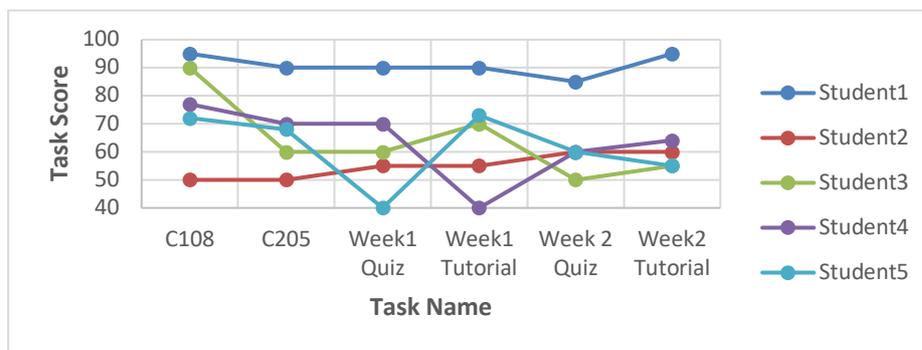

*Figure 4 - Example synthesised task score data*

### 4.2    Evaluation Metrics and Study Design

In this study, there are three quantitative metrics used to evaluate the responsiveness, adaptiveness, and effectiveness of our proposed methodology:

**Flesch-Kincaid Readability Score (FKRS):** assesses text complexity and readability, ranging from 0 to 100, based on sentence length and word syllable count [21]. This is crucial in higher education to tailor materials to students' comprehension, improving learning and performance. In computer science, domain-specific terms and coding may lower the score but are often not as challenging for students.

**Response Time:** is critical for providing feedback. Quick feedback keeps students engaged and helps them correct mistakes in real time, enhancing learning. Delays can cause frustration and disengagement.

**Specificity of Feedback (Feedback length):** involves detailed, tailored feedback that explains mistakes, reasons, and corrections. It also includes providing external sources like documentation, similar exercises, and GitHub repositories to enhance learning [22]. The number of relevant sentences is used to measure feedback length.

### 4.3    Results and interpretation

The pilot ITS has been tested across several metrics and benchmarks to evaluate the quality and relevance of the provided feedback. Three tutorial tasks were assigned to the system, and the feedback was assessed. The complexity of the tasks increases from Task 1 to Task 3.

The FKRS evaluated feedback across different student performance category levels as well as a general approach, without tailored prompt techniques. As shown in Figure 5, the overall scores output from the LLM are acceptable, even containing specific domain keywords and pieces of code. It can be observed that the readability score for a below-average student, a group that needs the higher readability scores the most, is higher than the non-contextual general feedback approach without using tailored prompts. FKRS scores in the 50 to 45 range are interpreted to be University-level readers, which is indicative of the higher complexity of the third task.

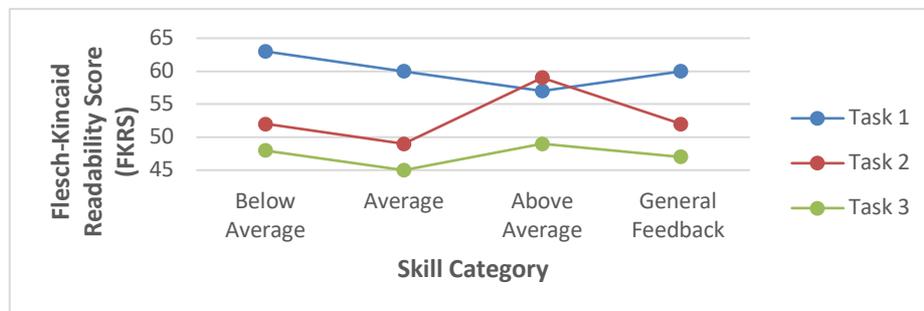

*Figure 5 - Flesch-Kincaid Readability Score by Skill Category*

Before applying the Self Consistency technique, 5 feedback responses are generated for each task. The final feedback is determined by a majority voting process. Therefore, the reported response time reflects the time required to generate 5 responses and select one based on the majority voting algorithm. Figure 6 illustrates that the time needed to generate final feedback is shorter for tailored feedback requests compared to general feedback requests. The reduced wait time for personalised feedback requests can be credited to the prompt engineering techniques used, which offer a step-by-step breakdown of the process for providing feedback. This enables quicker identification of the task's nature and faster feedback provision by the LLM. However, it is important to note that as the complexity of the task increases, the difference in response time between tailored and general feedback decreases. Additionally, the response time for below-average students is faster than for average, above-average and general feedback requests. As student performance improves, the response time also increases. This could be due to the need for more concise, high-level, and advanced recommendations for high-performing students.

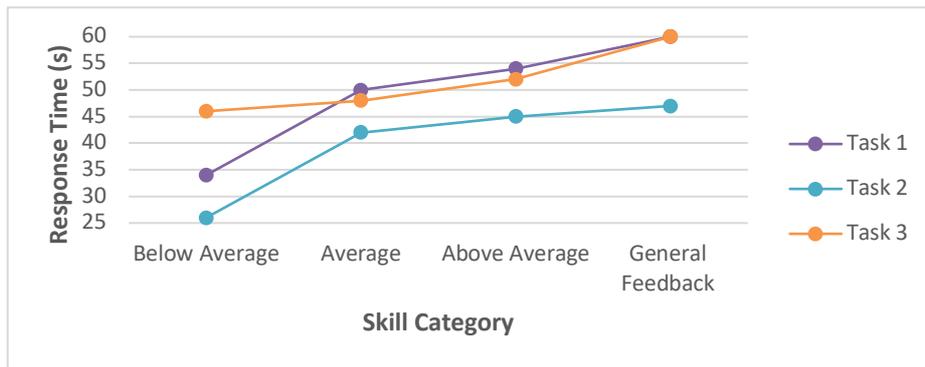

*Figure 6 - Response Time by Skill Category*

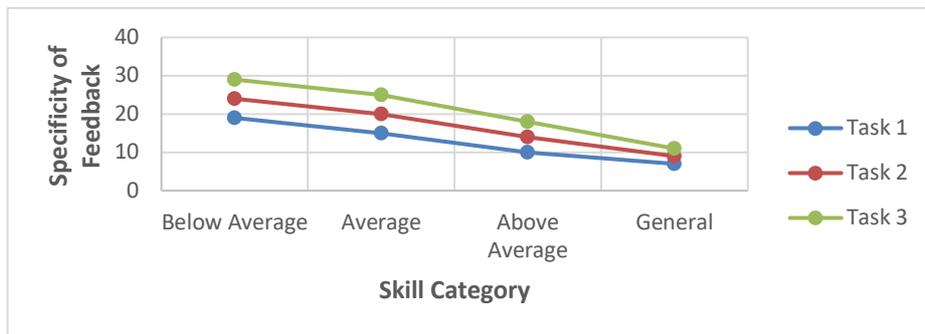

*Figure 7 - Specificity of Feedback by Skill Category*

When assessing the quality of feedback given to students with different performance levels, we can use the number of sentences, excluding code snippets, as metrics (Figure 7). These metrics give a better idea of how detailed and comprehensive the feedback is. For student performance categorised as below average, detailed feedback is provided with many sentences and explanations, along with extra tasks and resources for reinforcement as tasks become more complex. Average-performing students receive moderately detailed feedback, increasing in length with task complexity – instructions include combined steps and links to external resources for deeper understanding. Feedback for above-average performing students is concise, focusing on advanced techniques and assuming greater prior knowledge, with links to external resources for incorporating complex ideas. General feedback is brief and lacks guided structure, addressing only correcting mistakes without offering comprehensive step-by-step guidance, nor does it focus on the learning process or providing adaptive resources or detailed explanations, regardless of task complexity.

## 5     Conclusion and Future work

In conclusion, it is clear that providing tailored feedback in ITS can significantly improve student learning outcomes. The research used synthesised data to show that personalised feedback, tailored to individual student performance, has a positive impact on the learning process. The study's pilot system sorted students into three categories and offered context-aware, step-by-step guidance. This approach was found to be more effective in terms of FKRS, response time, and feedback specificity compared to general feedback methods.

One important recommendation for future work is to utilise actual student data to evaluate the proposed approach in a genuine educational environment. This method would offer a more precise and realistic assessment of the system's effectiveness, guaranteeing that the feedback given is truly tested and satisfied by students. Genuine data would assist in refining the system, making it more adaptable to the various learning requirements and situations of real students.

Furthermore, the study manually grouped the simulated students into three categories based on their skill levels and performance: *below average*, *average*, and *above average*. Although this method yielded promising results, future research could benefit from using more contextualised categories to classify students. This enhancement would provide a deeper understanding of student abilities and allow for more personalised feedback. Expanding the categorisation approach would help the ITS better cater to the specific needs of a wider range of students, ultimately improving the overall learning experience and educational outcomes.

**Disclosure of Interests.** The authors have no competing interests to declare that are relevant to the content of this article.